\documentclass{article}
\usepackage[utf8]{inputenc}

\usepackage[colorlinks=true,citecolor=blue,linkcolor=black]{hyperref}
\usepackage{graphicx}
\usepackage{amsmath}
\usepackage[version=4]{mhchem}
\usepackage{siunitx}
\usepackage{longtable,tabularx}
\usepackage{parskip}
\usepackage{times}
\usepackage{authblk}
\usepackage[top=1in, bottom=1in, left=1.1in, right=1.1in]{geometry}
\usepackage{titlesec}
\titleformat{\section}{\large\bfseries}{\thesection}{1em}{}
\usepackage{multirow}

\begin{document}
\title{The Probability of Vacuum Metastability and Artificial Vacuum Decay: Expert Survey Results\vspace{5mm}}

\author[1]{Jordan Stone}
\author[2]{Youssef Saleh}
\author[3]{Darryl Wright}
\author[4]{Jess Riedel}

\affil[1]{Department of Earth Science and Engineering, Imperial College London, UK}
\affil[2]{Department of Physics and Astronomy, University of Cairo, Egypt}
\affil[3]{Independent}
\affil[4]{Physics \& Informatics Laboratories, NTT Research, Inc., USA}

\date{}

\maketitle

\begin{abstract}
Vacuum decay posits that the universe's apparent vacuum is metastable and could transition to a lower-energy state. According to current physics models, if such a transition occurred in any location, a region of ``true vacuum" would propagate outward at near light speed, destroying the accessible universe as we know it by deeply altering the effective physical laws. Understanding whether advanced technology could potentially trigger such a transition has implications for existential risk assessment and the long-term trajectory of technological civilizations. We present results from what we believe to be the first structured survey of physics experts (n = 20) regarding both the theoretical possibility of vacuum decay and its potential technological inducibility. The average responded probability that our vacuum is metastable was 45.6\%, with an average estimated 18.8\% probability that arbitrarily advanced technology could induce vacuum decay if our vacuum is metastable. However, the survey revealed substantial disagreement among respondents on both whether the vacuum is metastable, and, if it is, whether vacuum decay could be artificially induced with arbitrarily advanced technology. According to participants, resolving these questions primarily depends on developing theories that go beyond the Standard Model of particle physics. Among respondents who considered vacuum decay theoretically possible, it was generally expected that artificial induction would pose significant technological challenges even for a civilization with galactic resources.

\end{abstract}
\vspace{5 mm}

\newpage

\section{Introduction}
Quantum field theory posits that the fundamental constituents of the universe are quantum fields permeating spacetime. The vacuum is the lowest-energy joint configuration of these fields and, by virtue of energy conservation, it is unchanging in time if it is left undisturbed. Particles emerge as quantized excitations of these fields, with their properties determined by the vacuum expectation values of the underlying fields. The Higgs field plays a pivotal role in endowing particles with mass through spontaneous symmetry breaking (Abokhalil 2023, Van Dam 2011).

Empirical investigations, notably at high-energy particle colliders, have substantiated the Standard Model's predictions, including the discovery of the Higgs boson (Atlas Collaboration 2012). These findings affirm the current vacuum configuration as a local minimum of the Higgs potential. However, theoretical extrapolations of the Standard Model suggest that this vacuum state may not be the absolute ground state (Markkanen et al. 2018). Instead, it could represent a metastable ``false vacuum," with a lower-energy ``true vacuum" existing elsewhere in the field configuration space (Espinosa et al. 2008). However, this conclusion relies on extrapolating the Standard Model to energy scales far beyond current experimental reach, so it must be viewed with caution.

The potential metastability of our vacuum raises the prospect of vacuum decay - a quantum tunnelling event wherein a region of space transitions to the true vacuum state (Markkanen et al. 2018). Such a transition would nucleate a bubble of true vacuum that expands at relativistic speeds, altering the fundamental constants and interactions within (Hut and Rees 1983). The implications of such an event are profound, potentially rendering the existing structures and laws of physics unrecognizable. 

Calculations indicate that the probability of spontaneous vacuum decay is exceedingly low on cosmological timescales (Isidori et al. 2001). However, local decay might occur if enough energy is concentrated in a small volume, or the fundamental fields are otherwise manipulated into a configuration that relaxes to the true vacuum rather than to our metastable vacuum (Tegmark and Bostrom 2005). No known natural energetic processes, even the largest supernovae, are remotely capable of this, but extremely advanced technology in the long-term future may change this. The mere possibility of vacuum decay may thus constrain the behavior, coordination, and longevity of advanced civilizations, influencing both the risks they accept and the limits of cosmic expansion. To explore these questions, we surveyed expert physicists for their views on (1) the likelihood our vacuum is metastable, and (2) the likelihood that if it is metastable, an arbitrarily advanced civilization could induce vacuum decay.

This paper presents findings from that survey. The conclusions do not aim to provide a view representative of the community, but to provide insights on arguments for and against the possibility of artificially induced vacuum decay.

\section{Approach}

In total, 105 researchers were identified with sufficient expertise in quantum field theory and cosmology to answer the survey. Researchers were identified by their authoring of peer reviewed literature on or related to vacuum decay and by evidence of expertise in relevant fields such as quantum field theory and early universe cosmology. 20 responses were received. Some physicists who declined to participate in the survey expressed concern that its results could be misinterpreted by the media as suggesting risks from current high-energy physics experiments. We emphasize that our team does not consider the survey findings to have any bearing on the safety of existing or planned physics facilities.

The survey addressed the metastability of the vacuum and the possibility of an artificial transition to a lower energy state. The key questions were:

\begin{itemize}
    \item ``How likely (0\%-100\%) do you think it is that the apparent vacuum of the quantum field theory describing our observable universe today is in fact only metastable, so that vacuum decay is physically possible?”
\end{itemize}
\begin{itemize}
    \item ``Conditional on the apparent vacuum being only metastable, how likely (0\%-100\%) do you think it is that a transition to a lower-energy vacuum could reliably be induced (purposefully or accidentally) with arbitrarily advanced technology?”
\end{itemize}

Respondents were then asked to explain their answers to each question and suggest research that would be needed to become more certain about their answer.

Respondents were also asked how confident they were in their responses, whether their opinion was informed by colleagues or their personal assessment, and what level of familiarity they have with the vacuum decay literature.

The anonymity of the respondents is kept, and respondents were unable to see each other's responses. 

\section{Results and Discussion}

\subsection{Respondents}

We received 20 responses from 17 institutions. 35\% of respondents were from the UK, 35\% from the USA, and 30\% from the rest of the World. The main backgrounds of respondents were in particle physics and quantum field theory (70\%) and cosmology (45\%). 85\% of respondents had published scientific articles on vacuum decay, and 50\% of respondents were professors (Figure 1)

\begin{figure}[h]
    \centering
    \includegraphics[width=0.7\textwidth]{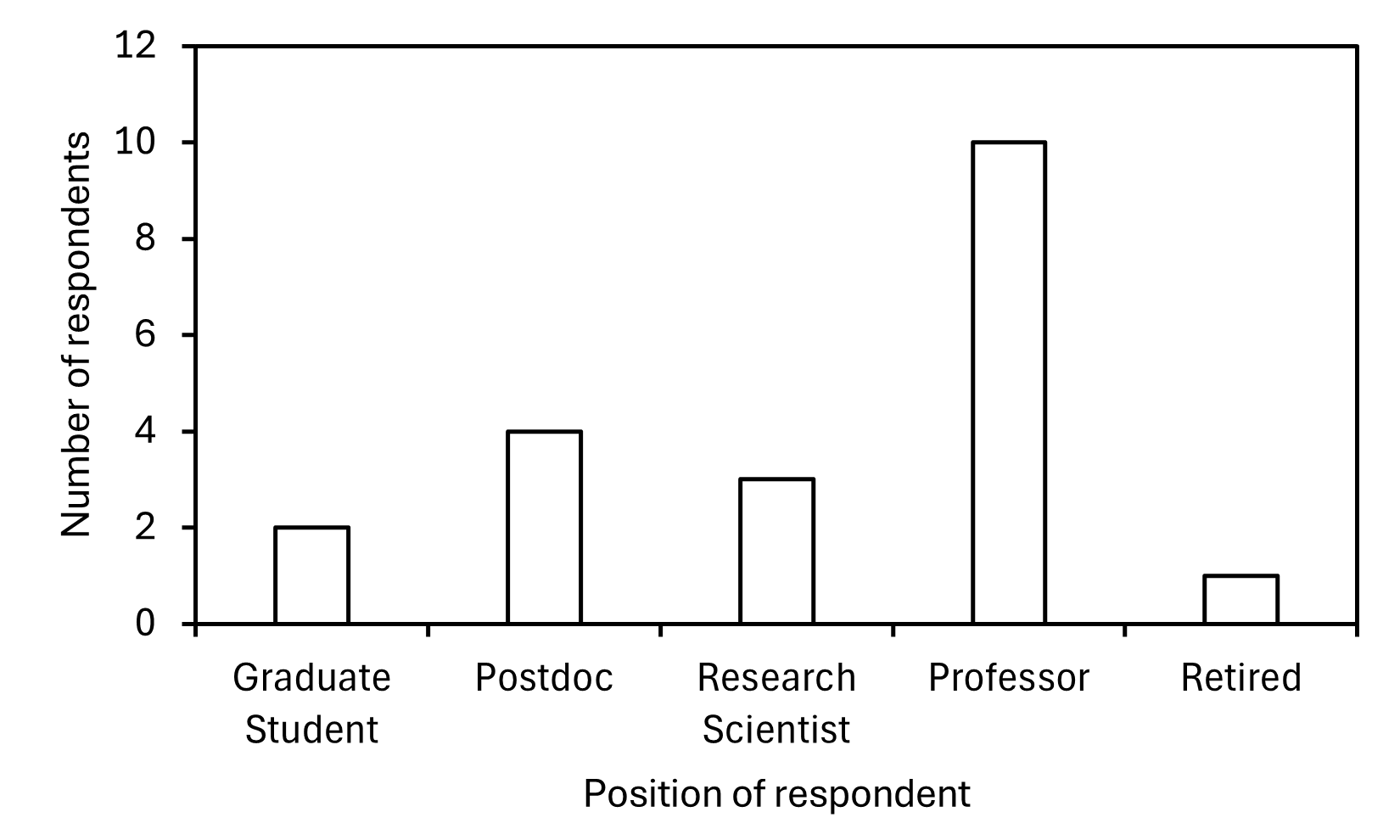}
    \caption{\small Positions of respondents to the survey. }
    \label{fig:example}
\end{figure}

At the end of the survey, respondents were asked, on a 1-5 scale, how confident they were that their estimates reflected the key scientific and philosophical considerations available at the present time (Figure 2a). Additionally, on another 1-5 scale, they were asked whether their opinion was mostly informed by the opinions of colleagues they trust (1) or their own personal assessment (5; Figure 2b). The mean responses to the questions of key considerations and own personal assessment were 4.0 and 3.8 out of 5, respectively. These results indicate that the responses likely reflect the considered opinions of experts in the field. 

\newpage
\begin{figure}[h]
    \centering
    \includegraphics[width=0.8\textwidth]{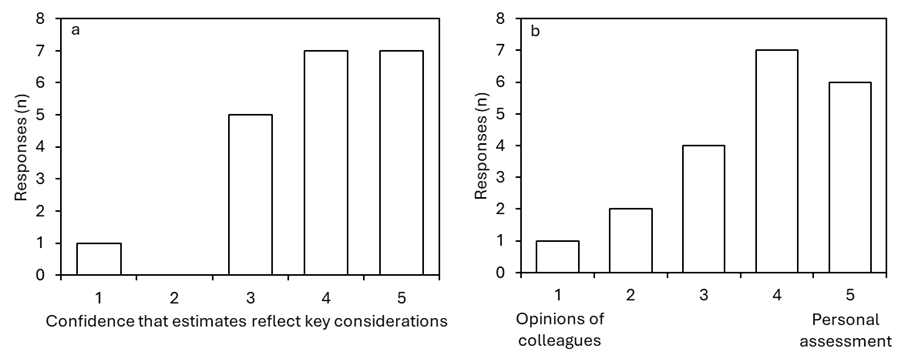}
    \caption{\small Respondents recorded what their opinions in the survey were based on. (A) Confidence that their estimates reflect the key scientific and philosophical considerations available at the present time. 1 = Low confidence, 5 = High confidence. (B) Response on whether their opinion on the possibility of vacuum decay is mostly informed by opinions of colleagues or personal assessment. 1 = opinions of colleagues, 5 = personal assessment. }
    \label{fig:example}
\end{figure}

\subsection{Is the Apparent Vacuum Metastable?}

Respondents were asked: ``How likely (0\%-100\%) do you think it is that the apparent vacuum of the quantum field theory describing our observable universe today is in fact only metastable, so that vacuum decay is physically possible?”. 

\begin{figure}[h]
    \centering
    \includegraphics[width=0.7\textwidth]{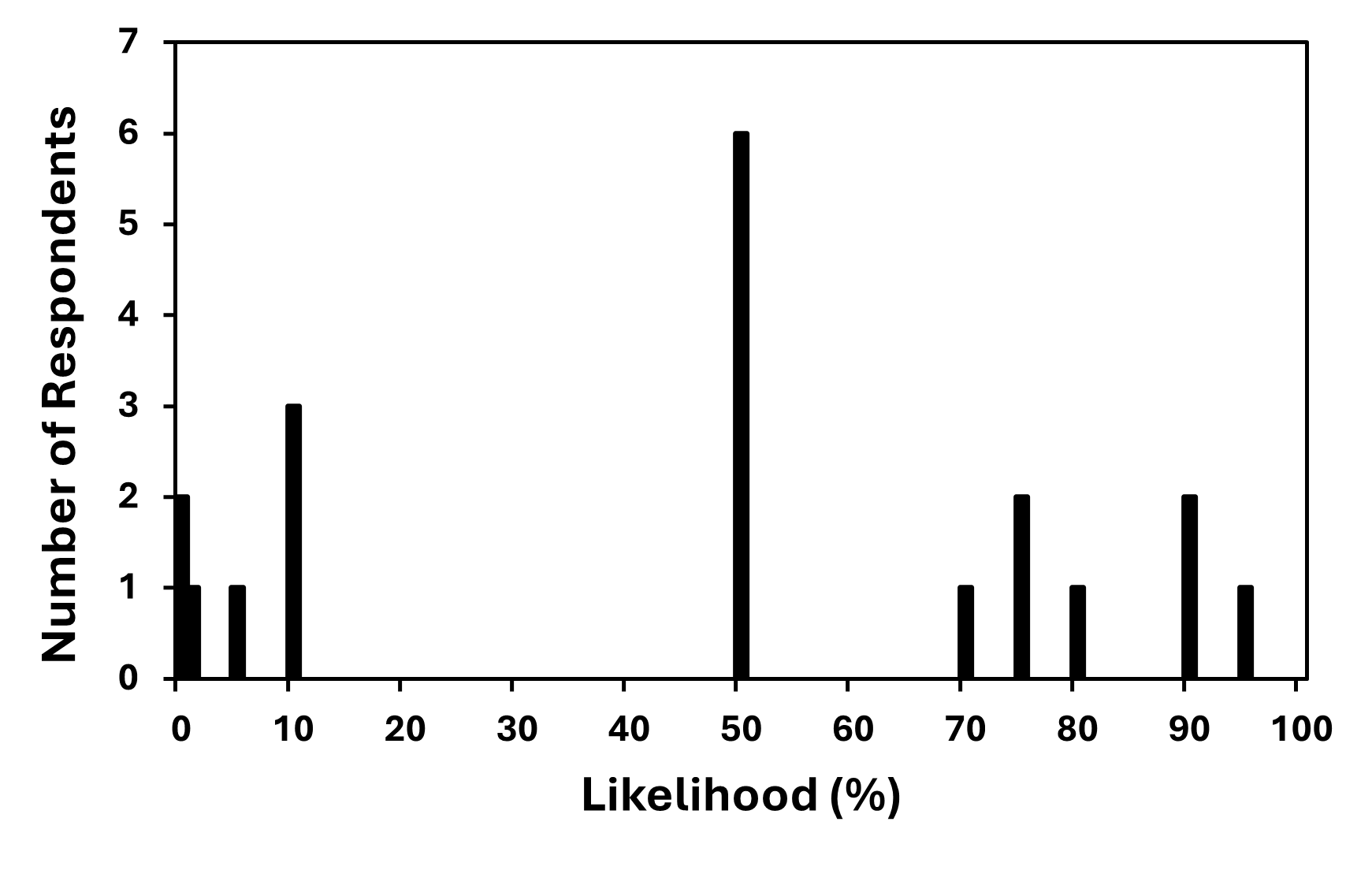}
    \caption{\small Responses to the question: ``How likely (0\%-100\%) do you think it is that the apparent vacuum of the quantum field theory describing our observable universe today is in fact only metastable, so that vacuum decay is physically possible?” The mean response was 45.6\%. n = 20, standard deviation = 33.7\%.}
    \label{fig:example}
\end{figure}

\vspace{3 mm}

The responses to this question were very mixed but could be divided into three groups based on their predicted percentage likelihood that the vacuum is metastable and the reasons given in the free-text explanation sections. Broadly, respondents fell into three stances: an agnostic group (50\%), a high-probability group (70–95\%), and a low-probability group (0–10\%).

One group of respondents (6 out of 20) selected 50\%, often indicating this as an ‘agnostic’ answer rather than a literal probability. In their free-text explanations, many said it is impossible to give a firm number because we cannot confidently extrapolate the Standard Model to the ultra-high energy scales relevant for vacuum decay. In other words, 50\% here meant they felt the likelihood was fundamentally unknowable given current physics.

A second subset (7 respondents) assigned a high likelihood (70–95\%) to metastability. Their rationale was that the Standard Model, given current measured parameters, explicitly predicts a metastable Higgs vacuum. Individual percentages varied, perhaps reflecting each person’s degree of trust in the Standard Model’s validity at energies beyond current experiments.

A ``low-probability" subset of respondents (7 of 20) selected 0-10\%. Many in this subset expressed the opinion that the Standard Model might be fundamentally wrong, particularly if its predictions are extended to extremely high energy scales (i.e. energy scales relevant to vacuum decay). Some felt that extending the Standard Model to scales relevant for vacuum decay is beyond its domain of validity, thus any prediction of metastability could be meaningless. In essence, this group suspects that new physics (not accounted for in the Standard Model) probably makes our vacuum stable. Note that one person responded 10\% as a geometric mean of widely divergent possibilities, reflecting their high uncertainty about physics at these scales, so more accurately fits in the 'agnostic' group. 

This division indicates that while many experts agree that metastability is predicted by present-day particle physics, there is a clear recognition of the Standard Model’s limitations, particularly in the context of vacuum stability. 4 of 7 comments on the ``additional comments” section indicated that it is extremely difficult to make a defined numerical prediction about the metastability of the vacuum. The uncertainty arises from the fact that the current theoretical framework may not fully capture the dynamics of the universe at very high energy levels or under extreme conditions. This suggests that deeper studies beyond the Standard Model would be needed to answer this fundamental question. 

14 of 20 explanations given by respondents refer to an agreement that the standard model predicts metastability (and so vacuum decay is possible), but the standard model is most likely wrong (or subject to change), especially at high energy levels (e.g., at the Planck scale). So we tentatively summarize the 3 groups as arising from the reasoning pathways presented in Figure 4.

\begin{figure}[h]
    \centering
    \includegraphics[width=0.8\textwidth]{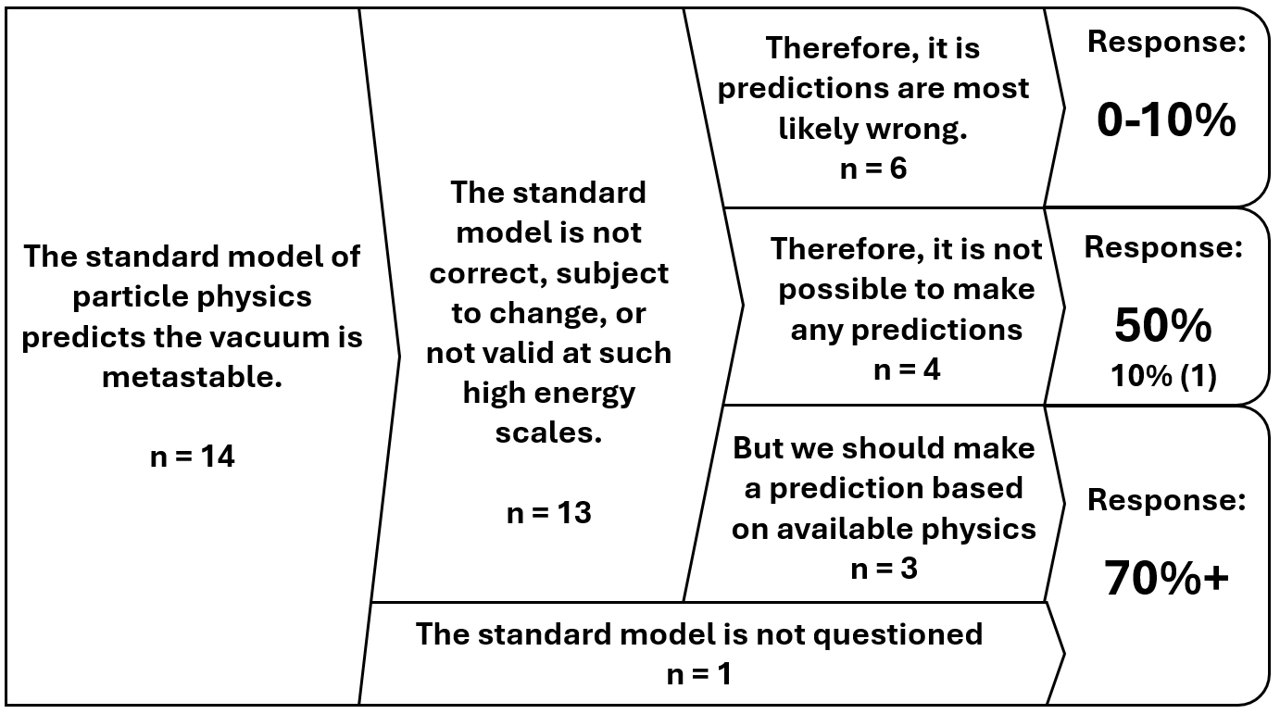}
    \caption{\small Reasoning pathways that explain most (14/20) of the responses to the question of whether our universe’s vacuum is metastable. One person responded 10\% as a geometric mean, so has been grouped with the ``50\%” responses. }
    \label{fig:example}
\end{figure}

The research that would best answer this question is therefore the development of physics beyond the Standard Model or further understanding of the validity of the Standard Model up to the Planck scale. This is at the core of the uncertainty behind the different reasoning pathways. 

Some additional reasoning was given that did not fit into Figure 4. This included the argument that most systems have multiple minima, so we are likely in a universe that is not at a minimum vacuum (n = 2, responding 50\% and 80\%). Another respondent argued that the quantum tunneling rate is likely high and responded 75\%. Three respondents indicated uncertainty on multiple aspects of the question or gave an unclear explanation, responding 50\%, 50\%, and 75\%.

In addition to these differing views, the survey respondents provided written comments on the areas of research they believe would help increase their certainty on the question. The most frequently mentioned research suggestion can be summarized as the development of theoretical physics beyond the Standard Model of particle physics (n=10), with a particular focus on high energy scales. Relatedly, some researchers suggested that precision measurements of the Higgs boson’s properties (n=4) would allow them to be more confident in their predictions. Other suggestions included researching early universe phase transitions and the high-energy cosmological history (n=4), as these may represent examples of the conditions under which the vacuum could shift. If phase transitions like vacuum decay are possible, they would likely have occurred during the high-energy conditions of the early universe. The search for new particles (n=3) was also suggested as an important pathway to uncover hidden aspects of the universe’s fundamental structure. Some respondents (n=2) proposed computational investigations into the vacuum decay process and its rate, while others (n=2) highlighted experimental phase transitions, such as those in liquid helium or particle coupling experiments, as potential areas of focus. Other suggestions (n=1) included research into the Hubble constant, the expansion rate of the universe, the multiverse (in relation to ``cosmic bubble collisions” and ``negative spatial curvature”), and analogous condensed matter systems.

In short, expert opinion on vacuum metastability is divided across the realm of probability. However, many experts remain agnostic, but suggest many directions to increase certainty on whether or not the universe's vacuum is metastable.

\subsection{Can Arbitrarily Advanced Technology Induce Vacuum Decay?}

The second key question asked respondents to assume the vacuum is metastable, and then estimate the probability that a transition to the true vacuum could be deliberately or accidentally induced by an arbitrarily advanced technology. Respondents were asked: ``Conditional on the apparent vacuum being only metastable, how likely (0\%-100\%) do you think it is that a transition to a lower-energy vacuum could reliably be induced (purposefully or accidentally) with arbitrarily advanced technology?”. The responses are presented in Figure 5.  

\begin{figure}[h]
    \centering
    \includegraphics[width=0.6\textwidth]{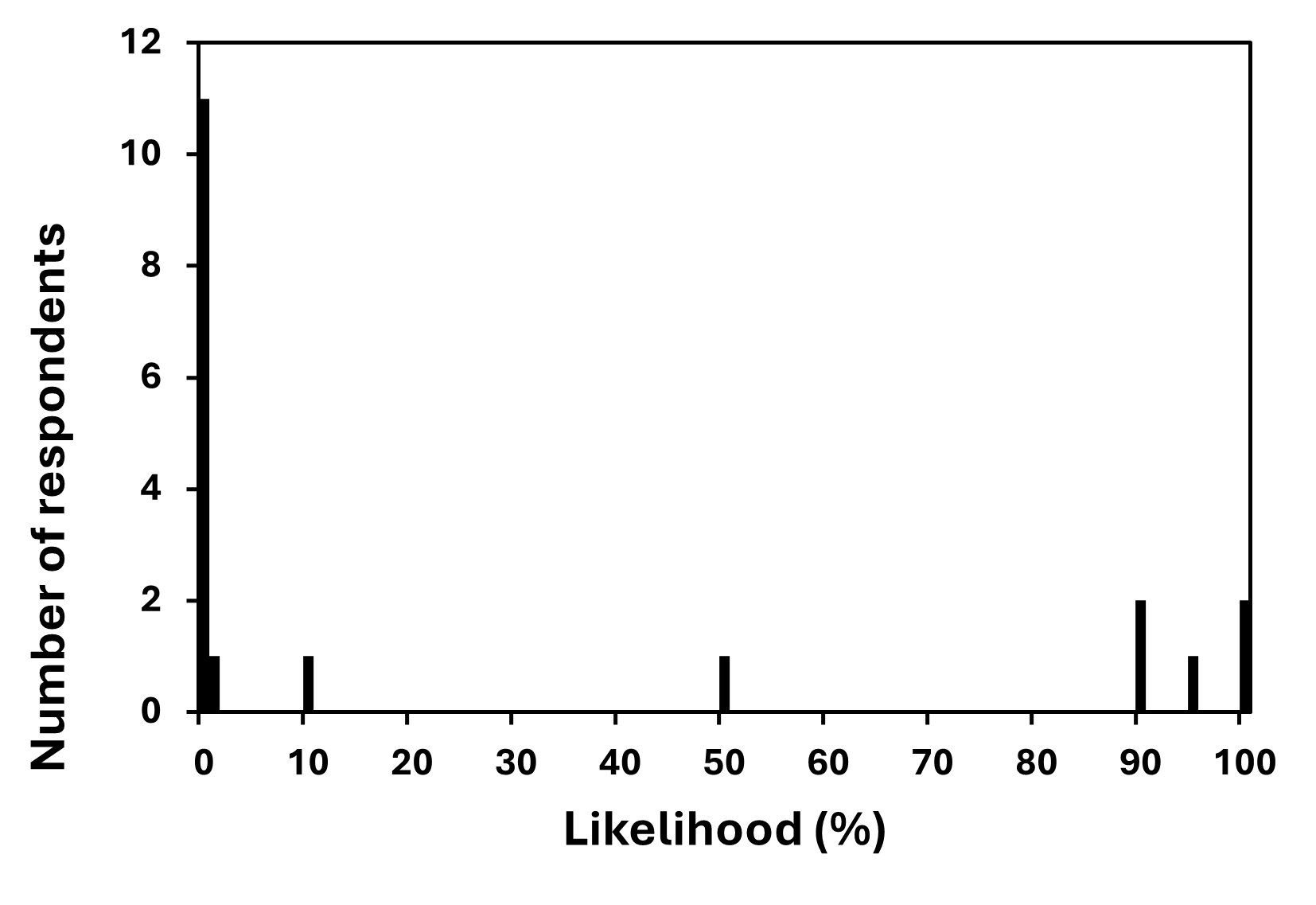}
    \caption{\small Respondents were asked ``Conditional on the apparent vacuum being only metastable, how likely (0\%-100\%) do you think it is that a transition to a lower-energy vacuum could reliably be induced (purposefully or accidentally) with arbitrarily advanced technology?”. The mean response was 18.8\%, n = 20, standard deviation = 42}
    \label{fig:example}
\end{figure}

In this case, responses were somewhat more convergent: 11 of 20 experts (55\%) answered that there is a 0\% likelihood that arbitrarily advanced technology could induce vacuum decay. There were a few different arguments given for responding 0\% to the probability of artificially vacuum decay. Firstly, vacuum decay would have occurred already in the long history of the universe if it was possible to induce it (n=4). Others (n=3) suggested that the energy scale of the instability required to induce vacuum decay is too high/quantum field theory does not allow it. Additionally, others (n=2) suggested that experiments would only act locally and not affect the Higgs vacuum (n=2).

The first argument referring to vacuum decay not having occurred in the universe's long history was presented from two angles. Firstly, the universe began in a hot and dense state (i.e., extremely high energy levels), so it seems that, if vacuum decay was possible to induce, then it would have happened then. Secondly, in the 13.7 billion year history of our universe, it’s likely that all the experiments we could theoretically execute would have already happened naturally. So there’s no technological way to conduct an experiment at a scale larger than anything that has already occurred.

However, all the responses from 90\% to 100\% (n=5) can be paraphrased by the statement: A sufficiently motivated civilization with arbitrarily advanced technology would be able to initiate vacuum decay because anything that does not violate the laws of physics would become possible. The respondents suggested some technologies that might be capable of inducing vacuum decay, such as solar-radius particle accelerators, Planck scale accelerators, and artificial black holes. The reason given by some for not responding “100\%” was because vacuum decay is highly theoretical, and it lacks any empirical evidence. So even if it is possible to induce, it might not be possible for us (i.e. physical beings accessing only 3 dimensions of this universe from this point in time and space) to induce, even with arbitrarily advanced technology.

Two respondents were in the middle.  The “10\%” respondent suggested that artificial black holes may induce false vacuum decay, and the “50\%” respondent stated that the question required pure speculation. 

The main research suggested to understand whether vacuum decay can be induced was research into artificially induced small scale transitions e.g. with liquid helium, which 4/11 respondents on this section suggested. Other suggestions made include: 
\begin{itemize}
    \item Theoretical physics, particularly Standard Model up to high energies (2 responses) 
\end{itemize}
\begin{itemize}
    \item Studies into the feasibility of building a giant accelerator in space (2 responses) 
\end{itemize}
\begin{itemize}
    \item Research into small-spatial-scale inhomogeneities (1 response) 
\end{itemize}
\begin{itemize}
    \item Research into false vacuum decay induced by black holes (1 response) 
\end{itemize}

Overall, compared to the previous question, there was more of a consensus, with 55\% of people responding that there is a 0\% chance that technologically induced vacuum decay is possible. Those opposed suggested that arbitrarily advanced technology implies that anything is possible if is allowed by the laws of physics – so, assuming vacuum decay is possible, then a civilisation with arbitrarily advanced technology would be able to induce it.

\subsection{Drivers of disagreement}

The debate surrounding vacuum stability doesn’t stem from a single source but from a complex mix of experimental, theoretical, and even philosophical factors. Respondents were allowed to select multiple reasons to explain the disagreement among experts. 50\% of respondents selected that interpreting experimental data, such as the precise mass of the Higgs boson, remains one of the greatest sources of uncertainty. Small changes in these measurements could dramatically alter our understanding of vacuum stability. 25\% of respondents selected unresolved mathematical challenges, particularly the lack of a complete theory that unifies quantum mechanics with gravity, which limits our ability to make definitive conclusions about the fate of the vacuum. 30\% of respondents selected philosophical perspectives like the conceptualization of stability, application of the anthropic principle, and the nature of the universe. An equal fraction of responses (30\%) suggest that this topic hasn’t yet received sufficient attention or thought, suggesting that vacuum stability, despite its profound implications, remains underexplored. These varied drivers highlight the challenge of answering such a fundamental question. 

Participants were asked to select which option drives most disagreement among experts (Figure 6). This lines up with the respondent’s comments discussed above that more experiments into small scale phase transitions are needed to understand whether technologically induced vacuum decay is possible. Though, developments in theoretical physics was toted as the main requirement to understand whether our vacuum is metastable. Though, it is possible that respondents also meant theoretical interpretation of experimental results along with other developments in theoretical physics such as addressing open mathematical questions.

\begin{figure}[h]
    \centering
    \includegraphics[width=0.9\textwidth]{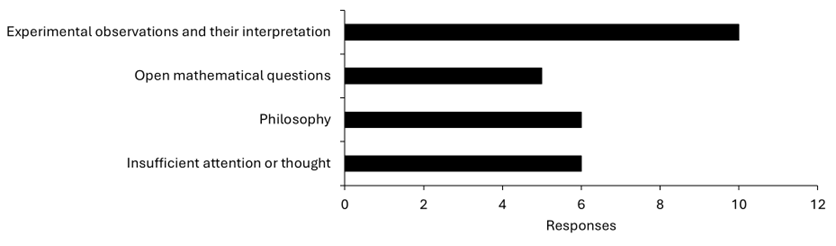}
    \caption{\small Perceived divers of disagreement among experts on this topic. Respondents could select multiple answers or choose none. }
    \label{fig:example}
\end{figure}

\subsection{Limitations}

The opposing arguments ``If vacuum decay was technologically inducible it would have already occurred naturally” and ``Anything not violating physical law is achievable with sufficiently advanced technology” potentially could be adjudicated by a more detailed account of the universe’s past and of the effective constraints any plausible technology would face. Our survey did not elicit the respondents’ position on these issues in sufficient details, which could be addressed in a future survey. 

Separately, recall that some people we emailed said they didn’t respond to the survey because they were concerned the results could be misrepresented by the media to suggest that current high energy physics experiments may be dangerous as they could induce vacuum decay. This could lead to backlash for the scientific community, especially people working on high energy physics. Through discussions with multiple physicists before undertaking the survey, we decided that the risk of this was low. However, we speculate that some people that did respond to the survey may have been motivated to underestimate the probability that technologically induced vacuum decay is possible based on the above concerns, most likely unconsciously due to past incidents. One person noted in the explanation for their answer that the survey ``just reminds me of people who claimed that experiments at CERN would create a black hole which would swallow the Earth”.
 
\section{Conclusion}
Understanding vacuum stability is not only a theoretical pursuit; it has potential implications for the longevity of the universe and the limits of advanced civilizations. While the risk of any near-term vacuum decay event is vanishingly small, exploring these expert views helps map the landscape of existential risks in the far future and identifies what scientific advances could clarify our cosmic situation. On average, our 20 experts estimated a ~45\% chance that our vacuum is metastable, but with opinions ranging from effectively 0\% to 95\%. Furthermore, even assuming metastability, more than half of these experts thought no technology could trigger vacuum decay, while a minority believed it would be feasible for a sufficiently advanced civilization (albeit very speculatively). According to respondents, resolving these questions primarily depends on developing theories that go beyond the standard model of particle physics, particularly relating to uncertainty on whether the Standard Model remains valid at the energy scales relevant to vacuum decay. Among respondents who considered vacuum decay theoretically possible, it was generally expected that artificial induction would pose significant technological challenges even for a civilisation with galactic resources. This work remains an exploratory survey of expert opinion, and further discussions and scientific advancements are required to reach consensus on vacuum decay. 

\section*{Acknowledgements}
We thank Carl Shulman and Adam Brown for inspiration and for feedback on the survey questions and write-up. We also thank the survey participants for sharing their expertise with us.

\section{Appendix A. Full survey}

\begin{figure}[h]
    \centering
    \includegraphics[width=1\textwidth]{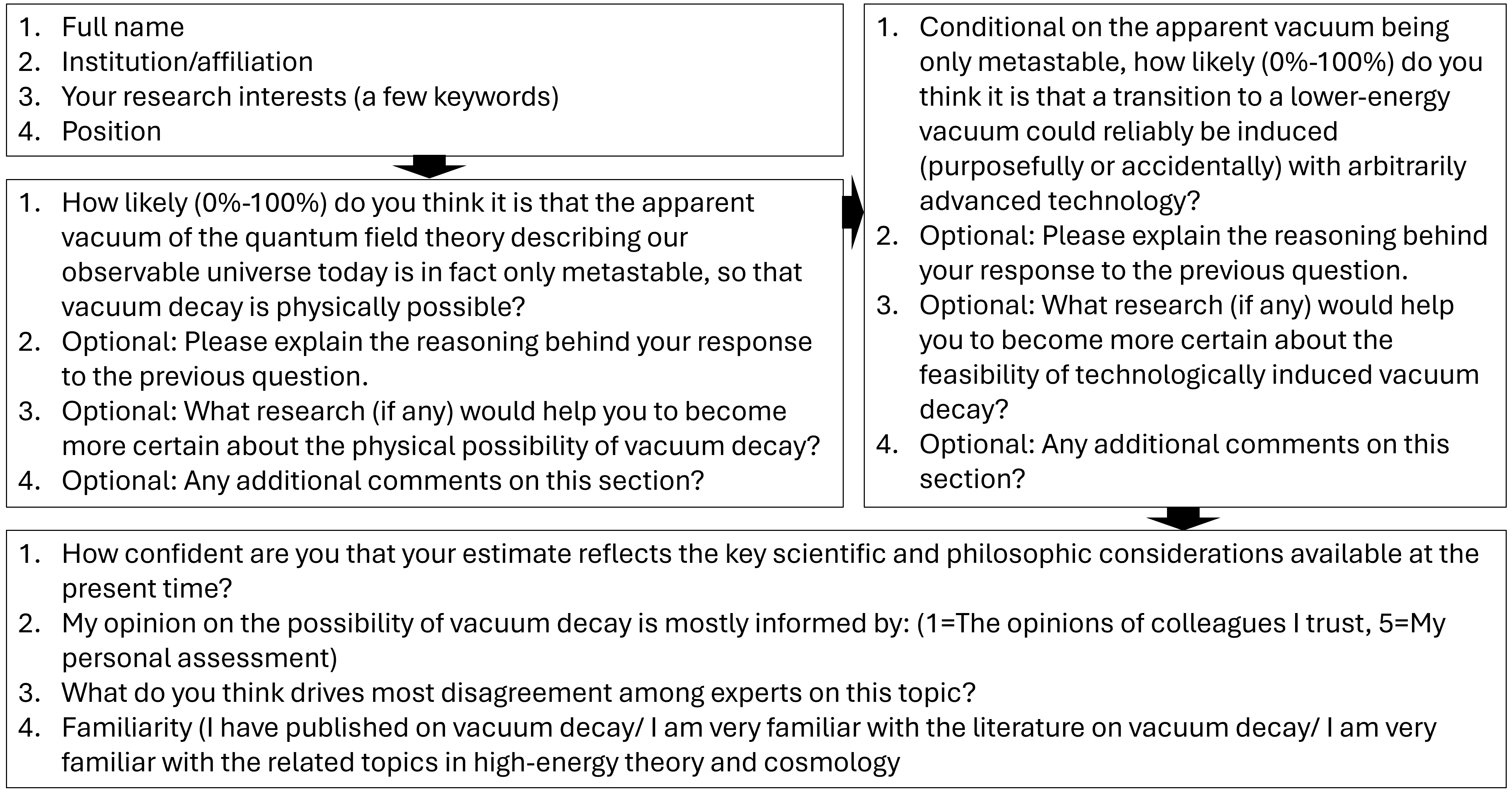}
    \caption{\small The full survey sent to experts in vacuum decay}
    \label{fig:example}
\end{figure}

\newpage
\section*{References}

Abokhalil, A. (2023). The Higgs Mechanism and Higgs Boson: Unveiling the Symmetry of the Universe. \textit{arXiv preprint} arXiv:2306.01019.

Atlas Collaboration (2012). Observation of a new particle in the search for the Standard Model Higgs boson with the ATLAS detector at the LHC. \textit{arXiv preprint} arXiv:1207.7214.

Espinosa, J., G. F. Giudice and A. Riotto (2008). Cosmological implications of the Higgs mass measurement. \textit{Journal of Cosmology and Astroparticle Physics} 2008(05): 002.
  
Hut, P., \& Rees, M. J. (1983). How stable is our vacuum?. \textit{Nature}, 302(5908), 508-509. 

Isidori, G., Ridolfi G. \& Strumia, A. (2001). On the metastability of the standard model vacuum. \textit{Nuclear Physics B} 609(3): 387-409.
 
Markkanen, T., Rajantie A., \& and Stopyra, S. (2018). Cosmological aspects of Higgs vacuum metastability. \textit{Frontiers in Astronomy and Space Sciences} 5: 40.

Tegmark, M. and N. Bostrom (2005). Is a doomsday catastrophe likely? \textit{Nature} 438(7069): 754-754.
  
\end{document}